\documentclass[utf8]{FrontiersinHarvard}

\usepackage{url,hyperref,lineno,microtype,subcaption}
\usepackage[onehalfspacing]{setspace}

\usepackage{amsmath}
\usepackage{amssymb}
\usepackage{mathtools}
\usepackage{amsthm}
\usepackage[capitalize,noabbrev]{cleveref}

\def\keyFont{\fontsize{8}{11}\helveticabold }
\def\firstAuthorLast{Maël Donoso}
\def\Authors{Maël Donoso\,$^{*}$}
\def\Address{$^{*}$Ouroboros Neurotechnologies, Lausanne, Switzerland}
\def\corrAuthor{Ouroboros Neurotechnologies, Place de la Riponne 5, 1005 Lausanne, Switzerland}
\def\corrEmail{mael.donoso@ouroboros-neurotechnologies.com}

\begin{document}
\onecolumn
\firstpage{1}

\title[A New Strategy for Artificial Intelligence]{A New Strategy for Artificial Intelligence: Training Foundation Models Directly on Human Brain Data} 

\author[\firstAuthorLast ]{\Authors}
\address{}
\correspondance{}
\extraAuth{}

\maketitle

\begin{abstract}

\section{}
While foundation models have achieved remarkable results across a diversity of domains, they still rely on human-generated data, such as text, as a fundamental source of knowledge. However, this data is ultimately the product of human brains, the filtered projection of a deeper neural complexity. In this paper, we explore a new strategy for artificial intelligence: moving beyond surface-level statistical regularities by training foundation models directly on human brain data. We hypothesize that neuroimaging data could open a window into elements of human cognition that are not accessible through observable actions, and argue that this additional knowledge could be used, alongside classical training data, to overcome some of the current limitations of foundation models. While previous research has demonstrated the possibility to train classical machine learning, deep learning, or reinforcement learning models on neural patterns, this path remains largely unexplored for high-level cognitive functions. Here, we classify the current limitations of foundation models, as well as the promising brain regions and cognitive processes that could be leveraged to address them, along four levels: perception, valuation, execution, and integration. Then, we propose two general methods that could be implemented to prioritize the use of limited neuroimaging data for strategically chosen, high-value steps in foundation model training: reinforcement learning from human brain (RLHB) and chain of thought from human brain (CoTHB). We also discuss the potential implications for agents, artificial general intelligence, and artificial superintelligence, as well as the ethical, social, and technical challenges and opportunities. We argue that brain-trained foundation models could represent a realistic and effective middle ground between continuing to scale current architectures and exploring alternative, neuroscience-inspired solutions. We also note that future discoveries in cognitive and computational neuroscience could make this strategy increasingly relevant over time, as new neural signals of interest are retroactively unlocked in present neuroimaging datasets. 
\tiny
 \keyFont{ \section{Keywords:} foundation models, human brain data, neuroimaging, brain-generated data, brain-trained foundation models, reinforcement learning from human brain (RLHB), chain of thought from human brain (CoTHB)}
\end{abstract}

\section{Introduction}

Foundation models, defined as models trained on broad data and adaptable to a wide range of downstream tasks, have rapidly emerged as the dominant paradigm in contemporary artificial intelligence (AI) \citep{bommasani_opportunities_2022}. The applications of foundation models are remarkably diverse, ranging from text \citep{brown_language_2020, touvron_llama_2023}, image \citep{ramesh_hierarchical_2022, alayrac_flamingo_2022}, and video \citep{singer_make--video_2022} generation to programming \citep{li_competition-level_2022}, robotics \citep{ahn_as_2022, driess_palm-e_2023}, mathematics \citep{collins_evaluating_2024}, and scientific research in domains such as physics \citep{farhadloo_towards_2025}, chemistry \citep{mendez-lucio_mole_2024}, and biology \citep{he_generalized_2025}. While these models can be trained on various data types, their most important source of knowledge is arguably the vast corpus of human-generated data, such as text and speech, which can serve as a semantic interface between different modalities \citep{ahn_as_2022, driess_palm-e_2023}. However, this data is ultimately the product of human brains, the filtered projection of a deeper, higher-dimensional neural complexity. In this theoretical paper, we explore the possibility of moving beyond human actions such as writing and speaking, and discuss how general-purpose foundation models could be trained directly on human brain data for improving their performance across a wide range of tasks. 

\begin{figure}[h]
\begin{center}
\includegraphics[width=17cm]{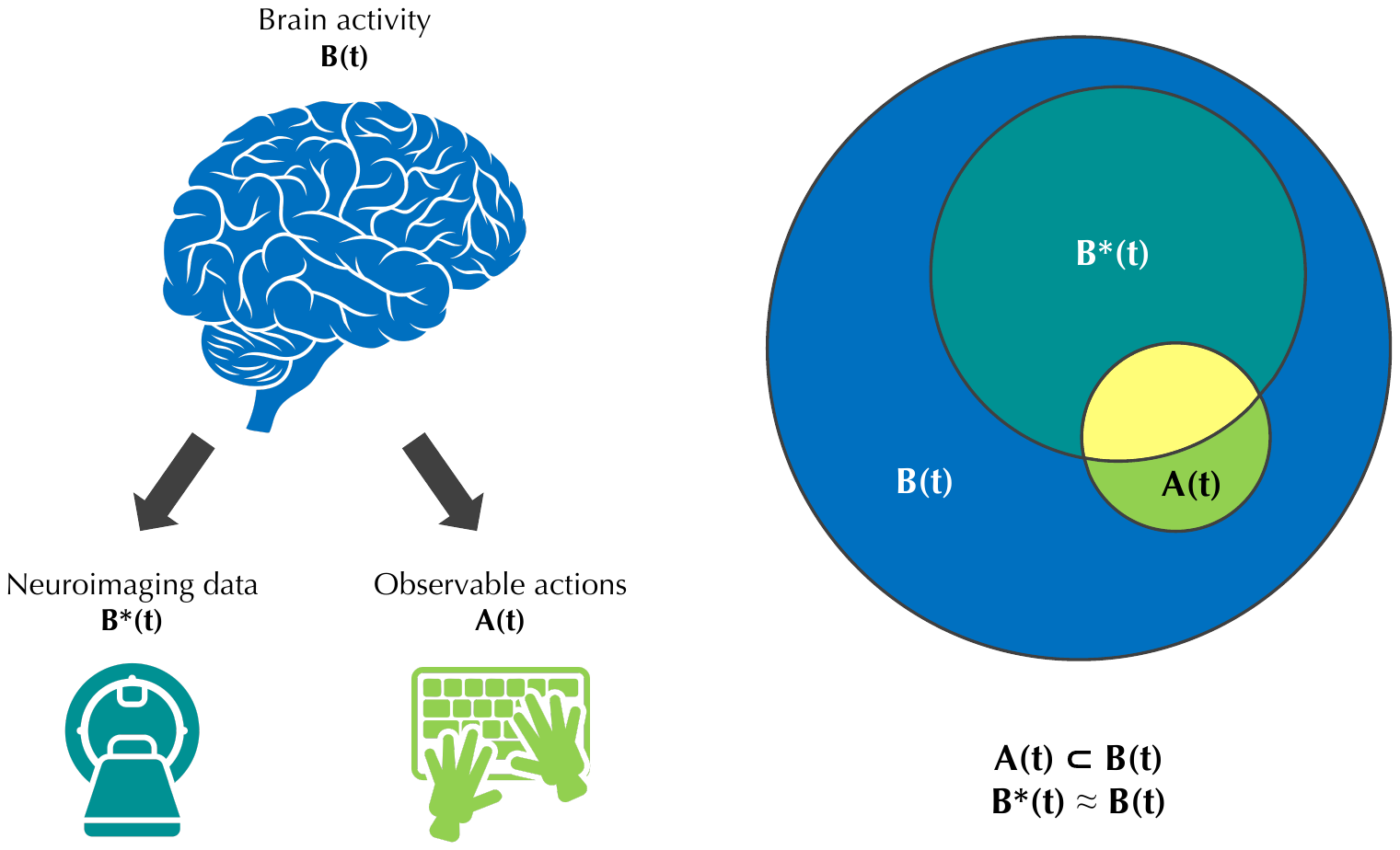}
\end{center}
\caption{\textbf{Generative brain principle. }\textit{Left:} Human brain activity, $B(t)$, can be approximated by neuroimaging data, $B^*(t)$, while generating observable actions, $A(t)$. \textit{Right:} The Venn diagram illustrates the relationship between the three sets, with the yellow area at the intersection of $B^*(t)$ and $A(t)$ representing the observable actions that can be decoded from neuroimaging data. The teal area, included in $B^*(t)$ but not in $A(t)$, represents the additional information that could be gained from incorporating neuroimaging data into foundation models. Future advances in neuroimaging technologies could gradually expand $B^*(t)$, resulting in a better approximation of $B(t)$.}\label{fig:generative_brain_principle}
\end{figure}

To formalize this idea in intuitive terms, let us introduce a simple notation. If $B(t)$ represents the brain activity of a human subject over time, and $A(t)$ the observable actions of this subject, then $A(t)$ could be considered a subset or projection of $B(t)$, since all voluntary actions are expected to originate in the brain activity. And if $B^*(t)$ represents the available neuroimaging data for the same subject, then $B^*(t)$ could serve as an approximation of $B(t)$. This deceptively simple idea, which we could call the \textit{generative brain principle}, has a surprisingly profound implication for foundation models, as illustrated in Figure~\ref{fig:generative_brain_principle}: it suggests that all human-generated data could, in principle, be directly inferred from human brain activity, within the limits of the quality and availability of neuroimaging data, and of the performance of brain decoding models. This theoretical upper bound could be further clarified with a short thought experiment. Let us consider the fact that natural language can already, to a certain extent, be decoded from neural signals, with models capable of reconstructing text from imagined \citep{tang_semantic_2023}, attempted \citep{willett_high-performance_2023}, or inner \citep{kunz_inner_2025} speech. Theoretically, this suggests that if we had access to unlimited neuroimaging data, we should be able to reconstruct any piece of text or speech ever included in the training corpus of a foundation model, up to a certain precision defined by the quality of the data and the performance of the decoding—an unrealistic, yet useful upper bound on the potential of human brain data for general-purpose foundation models. 

Practically, of course, it would make little sense to use the heterogeneous and expensive neuroimaging data at our disposal, $B^*(t)$, for the sole purpose of reconstructing text, speech, or other observable human actions, $A(t)$. It is obviously much more efficient to rely on the human-generated data already available, which is significantly simpler to process and more affordable to acquire. However, the true potential of human brain data is not to infer human actions, but to reveal the generative processes of human cognition underlying these actions. An emerging corpus of experimental research demonstrates that neuroimaging data can be leveraged to improve the performance of classical machine learning \citep{fong_using_2018}, deep learning \citep{spampinato_deep_2017}, or reinforcement learning \citep{iturrate_robot_2010} models on a variety of tasks, such as image recognition or robotic movement. Based on the discoveries in cognitive and computational neuroscience, we argue that a diversity of additional neural signals could potentially be leveraged to gain a deeper understanding of the processes that determine, for example, the choice of a particular word. While neuroimaging data provides only a simplified representation of the underlying neural complexity of the human brain, it could give us access to a cognitive latent space that partially includes, but vastly exceeds, the narrow bandwidth of observable behavior. In other words, we hypothesize that \textit{$B^*(t)$ could include elements of $B(t)$ that are not included in $A(t)$}, and that adding neuroimaging data to the training corpus of foundation models, alongside classical training data, could allow these models to move beyond surface-level statistical regularities, potentially overcoming some of their current limitations. Throughout this paper, expressions such as “training on human brain data” should be understood in their broadest sense, encompassing any form of learning, regularization, fine-tuning, or alignment that leverages neural signals in the training strategies of models. While a variety of specific implementations are possible, we argue that this general idea has the potential to serve as a new strategy for AI, and could represent the next frontier for improving foundation models. 

This would not be the first time that new possibilities emerge at the intersection of neuroscience and AI, as these two related domains share an intellectually rich and mutually beneficial history \citep{hassabis_neuroscience-inspired_2017}. Research on the human brain and other nervous systems has inspired the development of deep learning models such as multilayer perceptrons \citep{rumelhart_learning_1986}, convolutional neural networks \citep{lecun_gradient-based_1998}, and deep reinforcement learning agents \citep{mnih_human-level_2015}. In return, these models have significantly contributed to a better understanding of the human brain, by allowing researchers to decode complex neural patterns. However, recent advances in training classical machine learning, deep learning, or reinforcement learning models directly on human brain data open a particularly novel, and largely uncharted path, at the intersection of neuroscience and AI. It could be seen as a form of Copernican Revolution, where instead of designing models inspired by the human brain, researchers start with well-established models and guide them using neuroimaging data. While recent research has highlighted the potential of such methods, this path remains largely unexplored for high-level cognitive functions, and the implications of this approach for general-purpose foundation models have, to our knowledge, never been comprehensively investigated. In this paper, our objective is to develop a theoretical foundation for this important subject, highlighting a series of brain regions and cognitive processes whose neural signals could help overcome some of the current limitations of foundation models. We also propose two general methods that could be implemented to prioritize the use of limited neuroimaging data for strategically chosen, high-value steps in foundation model training, and discuss the potential long-term implications for agents, artificial general intelligence (AGI), and artificial superintelligence (ASI). 

\section{Foundation Model Training Data}

Foundation models are classically trained on a diversity of data types, which we could classify as either device-generated or human-generated. We use the term \textit{device-generated data} to refer to the signals that can be recorded without explicit human involvement or interpretation: for example, natural images, videos, sounds, chemical structures, genome sequences, or neuronal spikes. In the long term, the acquisition of such data should be primarily limited by the performance and deployment of technological devices. By contrast, we use the term \textit{human-generated data} to refer to the signals that are intentionally generated by humans, reflecting a form of semantic meaning or knowledge: for example, text, speech, code, or mathematical formulas, as well as the labels or descriptions associated with images, videos, or sounds. In the long term, the acquisition of such data should be primarily limited by the cognitive effort of human brains, which makes it more likely to be the limiting factor in the development of future foundation models \citep{shumailov_ai_2024}. Following our notation, human-generated data belongs to the set of observable human actions, $A(t)$, for any given subject. While large language models (LLMs) \citep{brown_language_2020, touvron_llama_2023} trained exclusively on human-generated data are already capable of extracting an impressive amount of factual \citep{petroni_language_2019} and commonsense \citep{sap_socialiqa_2019} knowledge, the combination of human-generated data with device-generated data has proved even more powerful, as demonstrated by the success of large multimodal models (LMMs) \citep{ramesh_hierarchical_2022, alayrac_flamingo_2022, singer_make--video_2022, ahn_as_2022, driess_palm-e_2023}, where human-generated data can serve as a semantic interface between different modalities of device-generated data. 

\begin{figure}[h]
\begin{center}
\includegraphics[width=17cm]{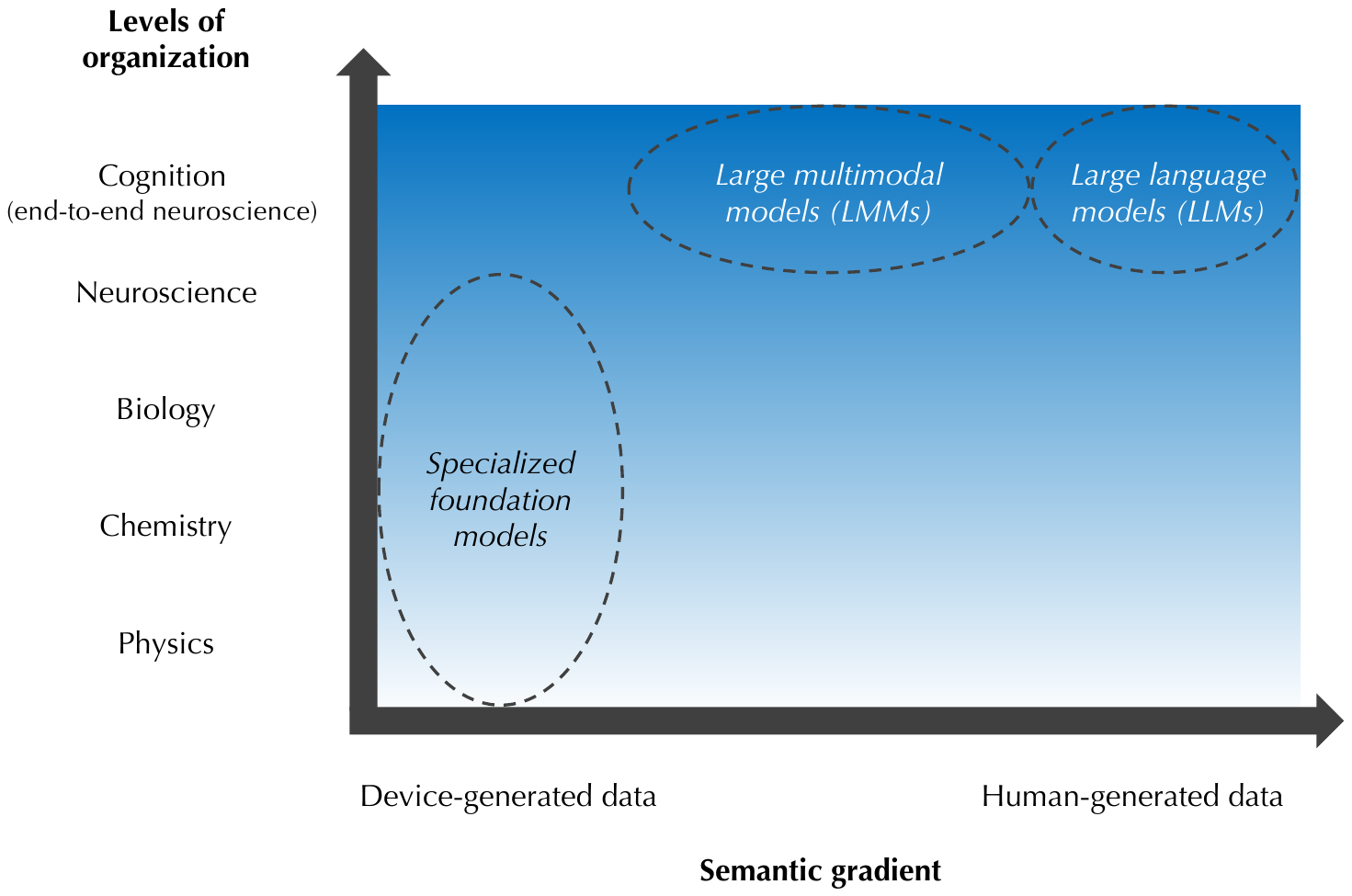}
\end{center}
\caption{\textbf{Foundation model continuum.} Foundation models are represented along two dimensions: the semantic gradient from predominantly device-generated data to predominantly human-generated data (x-axis), and the levels of organization (y-axis). Specialized foundation models are typically grounded in device-generated data, LMMs combine both data types, and LLMs rely primarily on human-generated data. Two empty regions highlight current constraints: the top-left corner, where it is currently difficult to imagine a foundation model of human cognition that does not significantly leverage human-generated data; and the bottom-right area, where it is equally difficult to imagine foundation models for natural sciences that do not primarily rely on device-generated data. Interestingly, a foundation model of human cognition grounded exclusively in neuroimaging data would naturally fall within the first empty region. However, we do not represent our hypothetical brain-trained foundation models there, since in the models we envision, neuroimaging data would serve as a complement, rather than a replacement, for more classical data types. The idea of fully brain-trained foundation models relying solely on neuroimaging data is intriguing, but beyond the scope of this paper.}\label{fig:foundation_model_continuum}
\end{figure}

To build a better understanding of the semantic gradient from device-generated data to human-generated data, let us consider the fact that some specialized foundation models are trained primarily or exclusively on recorded signals, combining for example natural videos with neuronal activity in mice \citep{wang_foundation_2025}. Even specialized AI systems, like those dedicated to the prediction of protein structure \citep{jumper_highly_2021} or the discovery of new crystals \citep{merchant_scaling_2023}, are occasionally considered to be foundation models, despite the fact that they are trained for a much narrower task than their general-purpose counterparts. With these particular cases at one end, we could visualize a \textit{foundation model continuum}, as illustrated in Figure~\ref{fig:foundation_model_continuum}: a progression from specialized foundation models, relying predominantly on device-generated data; to LMMs, trained on both classes of data; to LLMs, relying predominantly on human-generated data. Interestingly, the foundation models that exhibit the highest degree of generality, as well as the closest resemblance to human cognition, seem to stand at an intermediate position in this continuum. The training data of these particular models spans the full semantic gradient, including both natural inputs of the human brain, such as visual or auditory signals, and symbolic outputs of the human brain, such as text or speech. Yet even for these models, the human brain itself remains a black box. 

The foundation model continuum could be extended along another dimension: the classical series of levels of organization, where domains such as physics, chemistry, biology, and neuroscience build upon each other in a logical progression. Whereas specialized foundation models are often used to predict phenomena at the physical, chemical, or biological level, we suggest that general-purpose foundation models could be seen, with some caution, as end-to-end predictors for the neural level, since they are fundamentally trained to generate sensory or symbolic outputs based on corresponding inputs, in a way that approximates how a human brain would answer a request. Alternatively, these models could also be seen as predictors for the cognitive level, if we distinguish cognition as a domain of its own. Either way, this function sets them apart from the specialized foundation models developed specifically for neuroscience, sometimes known as brain foundation models \citep{donoso_human_2026}, which are typically not end-to-end but dedicated to the prediction of specific phenomena, such as neuronal spikes \citep{wang_foundation_2025}. Importantly, general-purpose foundation models, which will be our focus in the remainder of this paper, are not designed to emulate a particular human brain: rather, they seem to simulate an unrealistic, idealized form of human cognition, often combining considerable knowledge with a helpful attitude. Still, the possibility to formulate a request in natural language, and to receive an understandable answer that seems to be generated by a fellow human brain, is arguably the most important factor behind the ease of use and widespread adoption of LLMs and LMMs. Although seeing foundation models as end-to-end predictors for the neural level, or as predictors for the cognitive level, is admittedly an unusual perspective, it effectively highlights the oddity of considering the human brain as a black box. Paradoxically, even if LLMs and LMMs are implicitly designed to function as an idealized form of human cognition, their training strategies never leverage the insights of actual neuroimaging data. We argue that this missing piece could partially explain some of the current limitations of foundation models, which we discuss in the next section. 

\section{Foundation Model Architectures and Limitations}

Foundation models are typically based on the transformer architecture \citep{vaswani_attention_2017}, often combined with diffusion models \citep{ho_denoising_2020} for specific generative tasks. Their training strategies usually follow a three-step process: large-scale pre-training \citep{brown_language_2020} on a vast corpus of data to capture statistical regularities; supervised fine-tuning (SFT) \citep{ouyang_training_2022} on selected examples to optimize the model for its intended usage; and preference fine-tuning, typically conducted using reinforcement learning from human feedback (RLHF) \citep{christiano_deep_2017}, to align the model with human values. At inference time, techniques such as chain-of-thought (CoT) prompting \citep{wei_chain--thought_2023}, even with very simple prompts \citep{kojima_large_2022}, can improve the performance of foundation models when applied to complex tasks requiring step-by-step reasoning. While the combination of these approaches has allowed researchers and engineers to develop models of extraordinary capabilities, these methods remain dependent on observable human actions, $A(t)$, with no access to the generative processes of human cognition underlying these actions. For example, RLHF relies on human ratings, while CoT is guided using examples of human reasoning. Interestingly, the reliance on surface-level statistical regularities could partially explain some of the current limitations of foundation models, which we classify here along four levels. 

\subsection{Perception Level}

In the context of foundation models, we define \textit{perception} as the ability to construct robust representations of sensory inputs. Although multimodality tends to improve the robustness of foundation models \citep{radford_learning_2021}, these models remain surprisingly fragile in the way they represent perceptual information. Unlike the human brain, they are often sensitive to small perturbations or novel contexts, reducing their ability to generalize beyond their training data distribution \citep{hendrycks_many_2021}. One possible explanation for these fragile perceptual representations could be the fact that pre-training methods rely on human-generated data, such as the labels or descriptions associated with images, videos, or sounds, which only capture surface-level perceptual regularities and can drive the system toward superficial solutions \citep{geirhos_shortcut_2020}. Strategies to mitigate this effect include adversarial training \citep{madry_towards_2019} and data augmentation \citep{hendrycks_augmix_2020}, with the common objective of making the models more resistant to perturbations. 

\subsection{Valuation Level}

In the context of foundation models, we define \textit{valuation} as the ability to represent and align with human values. Although post-training methods such as RLHF are designed specifically for this objective, foundation models remain imperfectly aligned with human preferences. Unlike the human brain, they are often vulnerable to misalignment in novel or ambiguous contexts \citep{khamassi_strong_2024}. One possible explanation for this imperfect value alignment could be the fact that post-training methods rely on human-generated data, such as ratings, which only capture partial valuation processes and cannot easily represent human values \citep{casper_open_2023}. Strategies to mitigate this effect include extensions of RLHF, such as reinforcement learning from AI feedback, to optimize the use of human-generated data \citep{bai_constitutional_2022}, as well as new approaches, such as inverse reinforcement learning, to recover the implicit reward function of foundation models \citep{joselowitz_insights_2025}. 

\subsection{Execution Level}

In the context of foundation models, we define \textit{execution} as the ability to coordinate and sustain complex tasks. Although inference techniques such as CoT can improve their performance, foundation models remain limited in their reasoning and planning abilities. Unlike the human brain, they do not implement the equivalent of a robust working memory \citep{wang_unable_2025}, and struggle with reasoning and planning tasks outside their training data distribution \citep{stechly_chain_2025}. One possible explanation for this limited executive function could be the absence of mechanisms comparable to the cognitive control of the human prefrontal cortex \citep{russin_deep_2020}, and the difficulty of approximating such control from human-generated data alone. Strategies to mitigate this effect include extensions of CoT, such as tree of thoughts \citep{yao_tree_2023} or graph of thoughts \citep{besta_graph_2024}, which can expand the flexibility of foundation models at the expense of a greater computational cost. 

\subsection{Integration Level}

In the context of foundation models, we define \textit{integration} as the ability to form coherent concepts and generalize across novel domains. Although foundation models exhibit some elements typically associated with social cognition \citep{strachan_testing_2024}, their conceptualization of the world and its different agents often remains superficial. Unlike the human brain, they are prone to hallucinations, generating plausible yet wrong outputs, particularly in novel or ambiguous tasks \citep{huang_survey_2025}. One possible explanation for this shallow conceptual integration could be the absence of mechanisms comparable to the sensorimotor grounding of the human body \citep{xu_large_2025}, and the difficulty of approximating such grounding from human-generated data alone. Strategies to mitigate this effect include the development of methods to detect hallucinations \citep{farquhar_detecting_2024}, as well as the growing interest in new approaches inspired by biological brains \citep{marshall_are_2025}. 

Overall, these current limitations of foundation models seem to arise, at least partially, from the superficial nature of human-generated data. Strategies to overcome these limitations often involve scaling methods, such as augmenting pre-training data, expanding post-training feedback, or allocating more computational resources at inference time. However, beyond their increasing computational cost, it is unclear whether these methods will continue to yield significant progress in the long term, or whether they will reach a plateau. An alternative strategy may be the development of new foundation model architectures, perhaps inspired by biological brains. However, beyond the scientific complexity of the task, it is unclear whether such architectures would be able to extract significantly deeper regularities, assuming that they still rely on the same training corpus. Training foundation models directly on human brain data could circumvent these two difficulties, by relaxing the assumption that human-generated data is the only relevant resource for modeling human cognition. While the other strategies are focused on augmenting this resource or adapting the entire system to better accommodate its limited supply, brain-trained foundation models would take the different and complementary approach of adding neuroimaging data to their training corpus, in order to build a deeper representation of human cognition. Exploring the feasibility of this path requires evaluating the strengths and weaknesses of the different types of neuroimaging data at our disposal, which we review in the next section. 

\section{Human Brain Data}

Research in cognitive and computational neuroscience relies on a variety of neuroimaging techniques, each with specific strengths and weaknesses, as summarized in Table~\ref{tab:neuroimaging_techniques}. For example, electroencephalography (EEG) \citep{niedermeyer_electroencephalography_2005} measures the electrical activity of the brain, using electrodes positioned on the scalp. This technique is typically used to record the activity of cortical regions, with a low spatial resolution but a high temporal resolution. It is a relatively simple and inexpensive technique, compatible with wearable devices, and serves as the method of choice behind most commercial neural interfaces. By contrast, functional magnetic resonance imaging (fMRI) \citep{logothetis_what_2008} measures the metabolic activity of the brain indirectly, by detecting changes in the cerebral blood flow and oxygenation. This technique can be used to record the activity of both cortical and subcortical regions, with a high spatial resolution but a low temporal resolution. It is more complex and expensive, and requires the subject to remain immobile inside the scanner, but is widely used in research due to its detailed and comprehensive functional maps. Importantly, these two techniques are not mutually exclusive, since they can be combined through simultaneous EEG-fMRI acquisition \citep{lioi_simultaneous_2020}. Furthermore, the possibility to predict fMRI activity from EEG activity could eventually be leveraged to enhance neural interfaces, and potentially achieve near-fMRI precision while retaining the affordability and wearability of EEG devices \citep{li_neurobolt_2024, donoso_neural_2025}. 

\begin{table}
    \centering
    \renewcommand{\arraystretch}{1.3}
    \begin{tabular}{>{\centering\arraybackslash}p{4cm}
                >{\centering\arraybackslash}p{1.5cm}
                >{\centering\arraybackslash}p{1.5cm}
                >{\centering\arraybackslash}p{1.5cm}
                >{\centering\arraybackslash}p{3cm}
                >{\centering\arraybackslash}p{2.5cm}}
         & \textbf{EEG} & \textbf{fMRI} & \textbf{MEG} & \textbf{ECoG} & \textbf{fNIRS}\\
        \textbf{Spatial resolution} & Low & High & Medium & Very high & Low-medium\\
        \textbf{Temporal resolution} & High & Low & High & High & Medium\\
        \textbf{Brain coverage} & Cortex & Whole brain & Primarily cortex & Local cortex & Superficial cortex\\
        \textbf{Invasiveness} & No & No & No & Yes & No\\
        \textbf{Portability} & Yes & No & No & No (medical supervision) & Yes\\
        \textbf{Cost} & Low & High & High & High (surgery) / Low (recording) & Medium\\
    \end{tabular}
    \caption{\textbf{Neuroimaging techniques.} Five major neuroimaging techniques (EEG, fMRI, MEG, ECoG, fNIRS) are compared across six important dimensions (spatial resolution, temporal resolution, brain coverage, invasiveness, portability, cost). Each technique comes with specific strengths and weaknesses, potentially providing complementary paths for brain-trained foundation models.}
    \label{tab:neuroimaging_techniques}
\end{table}

Beyond EEG and fMRI, several other neuroimaging techniques are frequently used in cognitive and computational neuroscience. Magnetoencephalography (MEG) \citep{baillet_magnetoencephalography_2017} measures the magnetic fields generated by the electrical activity of the brain, using highly sensitive magnetometers. This technique improves the spatial resolution over EEG while maintaining its temporal resolution, at the expense of a greater complexity and cost. Electrocorticography (ECoG) \citep{parvizi_promises_2018} measures the electrical activity of the brain locally, using electrodes surgically implanted on the surface of the cortex. This technique provides both a high spatial resolution and a high temporal resolution, but its invasiveness currently limits its applications to medical contexts. Finally, functional near-infrared spectroscopy (fNIRS) \citep{ferrari_brief_2012} also offers interesting experimental possibilities, although restricted to superficial cortical regions. Like fMRI, this technique provides an indirect measure of the metabolic activity of the brain, but like EEG, it is a relatively simple and inexpensive method, compatible with wearable devices. Together, these neuroimaging techniques allow researchers to investigate human brain activity from several complementary angles, and can often be combined, either with each other, or with additional physiological and behavioral measures \citep{meissner_self-regulating_2024}. The possibility to predict one neuroimaging modality from another, a process which we could call neural translation, could further expand the corpus of human brain data at our disposal, and unlock new possibilities for understanding and leveraging neural complexity. 

Nevertheless, leveraging human brain data comes with specific challenges. In particular, neuroimaging data is often noisy \citep{liu_noise_2016} and distinctively individual \citep{finn_functional_2015}, requiring methods to ensure the alignment of neural signals across subjects \citep{haxby_common_2011}. These challenges are routinely addressed in cognitive and computational neuroscience, where the mitigation of noise and the normalization across subjects form the bedrock of virtually all current research. Still, they are worth noting here, since they remain a distinctive characteristic of human brain data. Furthermore, inferring a particular cognitive process from the activation of a particular brain region is not straightforward \citep{poldrack_can_2006}, although it remains a valid approach \citep{hutzler_reverse_2014} and methods can be developed to evaluate and improve its predictive power \citep{poldrack_inferring_2011}. The heterogeneity of neuroimaging datasets and their acquisition cost can also be challenging, but these problems are increasingly mitigated by the emergence of large-scale, standardized, open repositories for neuroimaging data \citep{markiewicz_openneuro_nodate}. Moreover, it seems reasonable to assume that the advances in brain-computer interfaces \citep{willett_high-performance_2023} will eventually lead to a wider adoption of wearable or implantable neural technologies, resulting in a significant increase in the volume of neuroimaging data. Overall, while these challenges are non-trivial, the convergence of multimodal acquisition, neural translation, open repositories, and neural interfaces is likely to make neuroimaging data increasingly compatible with the requirements of foundation model training. This could offer an unprecedented opportunity to incorporate human brain data into future foundation models, not as a mere subset of device-generated data, but as the approximation $B^*(t)$ of human brain activity $B(t)$, which is itself the superset of observable human actions $A(t)$ for any given subject. Therefore, alongside device-generated and human-generated data, we suggest that neuroimaging data could be distinguished as a new category, which we could call \textit{brain-generated data}, in the context of foundation model training. Evaluating the potential of such data for foundation models requires identifying which specific brain regions and cognitive processes could be targeted for neural signal extraction, a question we address in the next section. 

\section{Promising Brain Regions and Cognitive Processes}

\begin{figure}[h]
\begin{center}
\includegraphics[width=17cm]{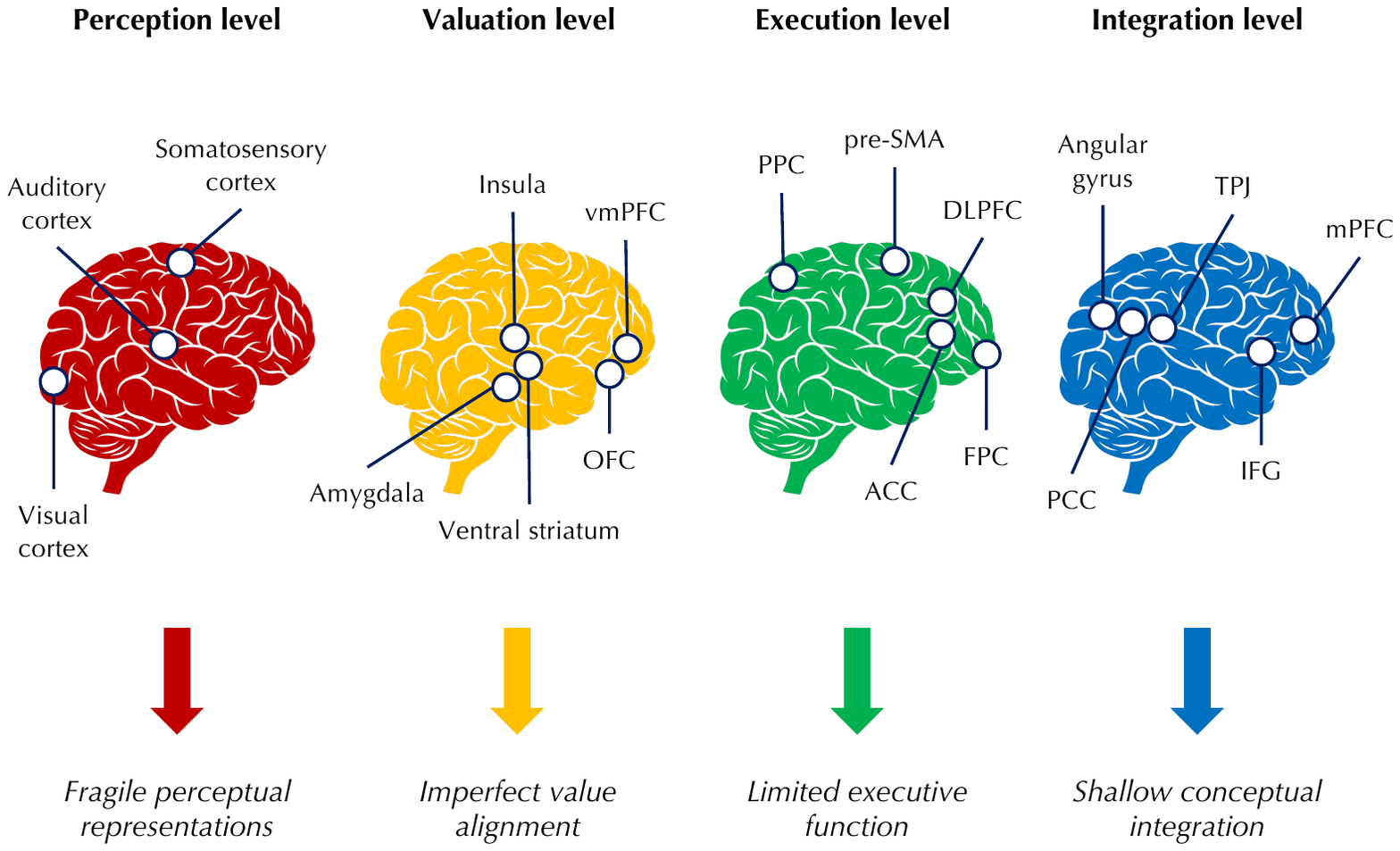}
\end{center}
\caption{\textbf{Cognitive levels.} \textit{From left to right:} Promising brain regions are schematically represented for the perception (red), valuation (orange), execution (green), and integration (blue) levels, along with the corresponding limitation of foundation models that such neural signals could help overcome. The positions of the brain regions are approximate, and provided only for illustrative purposes.}\label{fig:cognitive_levels}
\end{figure}

Brain-trained foundation models could potentially leverage a diversity of neural signals, associated with different brain regions and cognitive processes, as summarized in Table~\ref{tab:cognitive_levels} and illustrated in Figure~\ref{fig:cognitive_levels}. Interestingly, existing research on guiding classical machine learning, deep learning, or reinforcement learning models using neuroimaging data has primarily focused on visual \citep{fong_using_2018} and auditory \citep{freteault_alignment_2025} perception, or on reward and error valuation \citep{iturrate_robot_2010}. For perception, this choice was often motivated by the representational similarity between deep learning models trained for visual or auditory tasks and the corresponding sensory regions of the human brain, namely the visual cortex \citep{yamins_performance-optimized_2014} and the auditory cortex \citep{kell_task-optimized_2018}. For valuation, it was often motivated by the conceptual similarity between artificial and biological reinforcement learning \citep{sutton_reinforcement_2018}. Importantly, these foundational studies have been primarily conducted in the context of traditional applications of classical machine learning, deep learning, or reinforcement learning models, such as image recognition, audio classification, or robotic movement. However, the transition from narrow, specialized AI systems to broader, open-ended foundation models could represent a cognitive turning point regarding the most promising neural signals to consider. As the current limitations of foundation models extend beyond perception, and require AI research to move into the realm of valuation, execution, and integration, considering higher-order neural signals could become increasingly relevant, or even necessary. Here, we classify the most promising brain regions and cognitive processes for brain-trained foundation models along four levels, loosely organized by increasing degree of abstraction, and mapping to the current limitations of foundation models defined earlier. 

\begin{table}
    \centering
    \renewcommand{\arraystretch}{1.5}
    \begin{tabular}{>{\centering\arraybackslash}p{1.8cm}
                >{\centering\arraybackslash}p{2.8cm}
                >{\centering\arraybackslash}p{2.8cm}
                >{\centering\arraybackslash}p{4cm}
                >{\centering\arraybackslash}p{4cm}}
         & \textbf{Perception level} & \textbf{Valuation level} & \textbf{Execution level} & \textbf{Integration level}\\
        \textbf{Brain regions} & Visual cortex, auditory cortex, somatosensory cortex & Ventral striatum, ventromedial prefrontal cortex (vmPFC), orbitofrontal cortex (OFC), amygdala, insula & Dorsolateral prefrontal cortex (DLPFC), anterior cingulate cortex (ACC), frontopolar cortex (FPC), posterior parietal cortex (PPC), pre-supplementary motor area (pre-SMA) & Inferior frontal gyrus (IFG), angular gyrus, temporo-parietal junction (TPJ), medial prefrontal cortex (mPFC), posterior cingulate cortex (PCC)\\
        \textbf{Cognitive processes} & Vision, audition, touch & Reward, value, confidence, confirmation of a rule, novelty, salience, effort, self-control, internal state & Representation of current and alternative rules, evaluation of the environment, retrieval of a rule, motivation, attention, action selection & Language production and comprehension, conceptual integration, social cognition, self-cognition, autobiographical memory\\
        \textbf{Training approach} & Pattern decoding & Cognitive and computational inference & Cognitive and computational inference & Pattern decoding (language) / Cognitive and computational inference\\
        \textbf{Available research} & Medium & Low & Very low & Low\\
        \textbf{Foundation model limitation} & Fragile perceptual representations & Imperfect value alignment & Limited executive function & Shallow conceptual integration\\
    \end{tabular}
    \caption{\textbf{Cognitive levels.} The four cognitive levels (perception, valuation, execution, integration) are compared across five important dimensions (brain regions, cognitive processes, training approach, available research, foundation model limitation). For each level, the table highlights a selection of promising brain regions and cognitive processes whose neural signals could help overcome the corresponding limitation of foundation models.}
    \label{tab:cognitive_levels}
\end{table}

\subsection{Perception Level}

\textit{Perception} engages various regions of the human brain, among which the visual cortex \citep{dicarlo_how_2012}, the auditory cortex \citep{bizley_what_2013}, and the somatosensory cortex \citep{kaas_evolution_2008}. Together, these brain regions acquire and process primary sensory information, which is then contextualized into secondary perceptual representations. Foundational research has demonstrated that fMRI data can be leveraged to improve classical machine learning models for an object recognition task, while also suggesting that such methods could be extended to other neuroimaging techniques, learning algorithms, and sensory modalities \citep{fong_using_2018}. Two other foundational studies have shown that EEG data can be used to guide deep learning models, whether for an object recognition task \citep{spampinato_deep_2017} or an image generation task \citep{palazzo_generative_2017}, while similarly hinting at a possible extension of such methods beyond the visual modality \citep{spampinato_deep_2017}. Recent research has continued to explore the opportunity to guide deep learning models for visual tasks using fMRI data \citep{lu_teaching_2025, nguyen_brainformer_2025, lang_neurofusionnet_2025, zhao_neuralood_2025} and EEG data \citep{palazzo_decoding_2021, hansen_brain-guided_2025}, while also extending the investigation to auditory tasks using fMRI data \citep{freteault_alignment_2025, moussa_improving_2025}. The methods implemented in these studies are relatively variable, suggesting that a diversity of approaches can be considered to improve classical machine learning or deep learning models with neuroimaging data. Nevertheless, the training strategies usually involve some form of model regularization, fine-tuning, or another constraint based on neural signals, and they often leverage existing models pre-trained for visual or auditory tasks. Interestingly, rather than acquiring a new neuroimaging dataset for each experiment, it is sometimes possible to rely on model-generated data, using for example a pre-trained fMRI prediction model to generate large-scale synthetic fMRI data without additional cost \citep{zhao_neuralood_2025}. These results suggest that the brain regions and cognitive processes associated with the perception level could be promising targets to overcome the fragile perceptual representations observed in foundation models, at least if such methods can be scaled. We speculate that guiding robotics models using neural signals from the somatosensory cortex may also be an interesting opportunity to explore. 

\subsection{Valuation Level}

\textit{Valuation} engages a network of cortical and subcortical regions across the human brain. Among them, the ventral striatum is associated with reward prediction \citep{odoherty_dissociable_2004}, including when the task involves planning \citep{daw_model-based_2011}. This subcortical structure serves as a motivational node even in subliminal settings \citep{pessiglione_how_2007}, driving both cognitive and motor effort \citep{schmidt_neural_2012}, as well as novelty-seeking behavior \citep{wittmann_striatal_2008}. In complex reasoning tasks, the ventral striatum also encodes the confirmation of a new rule \citep{donoso_foundations_2014}. Another important region is the ventromedial prefrontal cortex (vmPFC), which serves as a general valuation node encoding the value of the elements present in the environment \citep{lebreton_automatic_2009}, while a distinct network encodes the effort necessary to acquire such elements \citep{prevost_separate_2010}. The valuation processes of the vmPFC are conducted independently of the category of goods, suggesting the idea of a common currency \citep{chib_evidence_2009}, and they are modulated by the degree of self-control exerted by the subject \citep{hare_self-control_2009}. The vmPFC is also engaged in more abstract valuation processes, encoding the value of the chosen action \citep{wunderlich_neural_2009}, the value of the current rule \citep{kolling_neural_2012}, and even the confidence associated with this rule \citep{de_martino_confidence_2013}. Meanwhile, a closely related brain region, the orbitofrontal cortex (OFC), is associated with the magnitude of the rewards obtained \citep{odoherty_abstract_2001} and the willingness to pay in economic transactions \citep{plassmann_orbitofrontal_2007}, while being flexibly regulated according to the current state of the subject \citep{gottfried_encoding_2003}. Other important brain regions for valuation include the amygdala, which serves as an emotional node particularly involved in salience mechanisms \citep{phelps_contributions_2005}, and the insula, associated with interoception and a variety of affective processes \citep{bud_craig_how_2009}. Together, these brain regions process rewards, assign a value to perceptions, actions, or rules, and motivate decision-making based on this valuation. 

Although several studies have demonstrated that reward and error signals extracted from EEG data can be used to guide reinforcement learning models in robotics \citep{iturrate_robot_2010, kim_intrinsic_2017, xu_accelerating_2020, kar_eeg-induced_2022}, research on using higher-level valuation processes for this objective appears to be limited. Here, we see a clear opportunity to extend the set of neural signals at our disposal, in order to leverage the full range of valuation processes of the human brain. For example, using fMRI to record simultaneously the activity of the ventral striatum, vmPFC, OFC, amygdala, and insula could provide us with a comprehensive monitoring of the valuation level. This monitoring would range from rewards \citep{odoherty_abstract_2001, odoherty_dissociable_2004, daw_model-based_2011}, to values \citep{plassmann_orbitofrontal_2007, lebreton_automatic_2009, chib_evidence_2009, wunderlich_neural_2009, kolling_neural_2012}, to confidence \citep{de_martino_confidence_2013}, to the confirmation of a valuable rule \citep{donoso_foundations_2014}. It would also allow us to control for related variables such as novelty \citep{wittmann_striatal_2008}, salience \citep{phelps_contributions_2005}, effort \citep{prevost_separate_2010, schmidt_neural_2012}, self-control \citep{hare_self-control_2009}, and the internal state of the subject \citep{gottfried_encoding_2003, bud_craig_how_2009}. Critically, such neural signals could extend far beyond the preference ratings currently used to align foundation models, opening a window into the generative processes underlying the expression of human values. Furthermore, at least some of these processes are automatic, requiring no voluntary action or intent \citep{pessiglione_how_2007}, thus offering a degree of objectivity and immediacy that distinguishes brain-generated data from human-generated data. Therefore, the brain regions and cognitive processes associated with the valuation level could be promising targets to overcome the imperfect value alignment observed in foundation models, provided that these neural signals can be effectively integrated into the training strategies of the models. 

\subsection{Execution Level}

\textit{Execution} engages a network of cortical and subcortical regions across the human brain, with the prefrontal cortex playing a central role. Among these regions, the dorsolateral prefrontal cortex (DLPFC) is associated with the selection of representations in working memory \citep{rowe_prefrontal_2000}, and serves as a general node for the implementation of cognitive control \citep{macdonald_dissociating_2000}. The architecture of this prefrontal region reflects a cascade of increasingly abstract executive processes, capable of controlling behavior based on stimuli, context, and instructions \citep{koechlin_architecture_2003}, with each level driven by dedicated motivational mechanisms \citep{kouneiher_motivation_2009}. This organization along several levels of abstraction is consistent with the behavior of subjects with frontal damage \citep{badre_hierarchical_2009}, and with research on the acquisition of abstract rules \citep{badre_frontal_2010}. In complex reasoning tasks, the DLPFC also encodes the retrieval of a previous rule whose reliability continues to be monitored in working memory \citep{donoso_foundations_2014}. Another important region is the anterior cingulate cortex (ACC), which encodes the volatility of the environment \citep{behrens_learning_2007} and serves as a general node for evaluating the outcome of actions, integrating rewards, errors, and conflict \citep{alexander_medial_2011}. In exploration-exploitation trade-off settings, the ACC represents the value of the exploratory behavior \citep{kolling_neural_2012}. Meanwhile, another region associated with exploration, the frontopolar cortex (FPC), encodes the reliabilities of alternative rules \citep{boorman_how_2009}, which are continually updated based on the feedback from the environment \citep{boorman_counterfactual_2011}, with current and alternative rules driven by dedicated motivational mechanisms \citep{charron_divided_2010}. Other important brain regions for execution include the posterior parietal cortex (PPC), which serves as an attentional node particularly involved in sensorimotor transformation \citep{andersen_intention_2009}, and the pre-supplementary motor area (pre-SMA), associated with different types of complex action selection \citep{nachev_functional_2008}. Together, these brain regions process the outcome of actions, assign a reliability to current and alternative rules, and implement decision-making based on stimuli, context, and instructions. 

Intriguingly, despite the interest in replicating the functions of the human prefrontal cortex in AI systems \citep{russin_deep_2020}, and although neural signals from brain regions such as the DLPFC have been used in neurofeedback studies \citep{godet_functional_2024}, research on training models directly on these regions appears to be limited, with the exception of a recent study in which invasive recordings from the cingulate cortex, possibly including the ACC, have been used to guide a model for a decision-making task \citep{aquino_brain2model_2025}. Here, as in the valuation level, we see a clear opportunity to extend the set of neural signals at our disposal, in order to leverage the full range of execution processes of the human brain. For example, using fMRI to record simultaneously the activity of the DLPFC, ACC, FPC, PPC, and pre-SMA could provide us with a comprehensive monitoring of the execution level. This monitoring would range from the representation and implementation of rules \citep{rowe_prefrontal_2000, macdonald_dissociating_2000, koechlin_architecture_2003, badre_hierarchical_2009, badre_frontal_2010}, to the evaluation of the environment \citep{behrens_learning_2007, alexander_medial_2011, kolling_neural_2012}, to the representation of alternative rules \citep{boorman_how_2009, boorman_counterfactual_2011}, to the retrieval of a previous rule \citep{donoso_foundations_2014}. It would also allow us to control for related variables such as motivation \citep{kouneiher_motivation_2009, charron_divided_2010}, attention \citep{andersen_intention_2009}, and action selection \citep{nachev_functional_2008}. Critically, such neural signals could extend far beyond the reasoning examples currently used to guide foundation models for multi-step tasks, opening a window into the generative processes underlying the decision steps of human reasoning. The concept of rule, or more appropriately task set \citep{sakai_task_2008}, defined as a configuration of cognitive processes that is actively maintained for executing a task, could provide an effective framework for interpreting neuroimaging data in this context. Therefore, the brain regions and cognitive processes associated with the execution level could be promising targets to overcome the limited executive function observed in foundation models, provided, as in the valuation level, that these neural signals can be effectively integrated into the training strategies of the models. 

\subsection{Integration Level}

\textit{Integration} engages an intricate network of cortical and subcortical regions across the human brain, with several regions playing prominent roles. Among them, the inferior frontal gyrus (IFG) serves as a general node for language production and comprehension \citep{friederici_brain_2011}, and the angular gyrus for conceptual integration \citep{binder_where_2009}. The temporo-parietal junction (TPJ) is particularly involved in social cognition \citep{carter_nexus_2013}, while the medial prefrontal cortex (mPFC) and posterior cingulate cortex (PCC) are important nodes of the default network, supporting self-cognition and autobiographical memory \citep{buckner_brains_2008}. Together, these brain regions allow humans to engage in symbolic thought and expression, understand the intentions of others, and reason about themselves. The complexity of these cognitive functions, and their distributed and intricate nature, make the integration level difficult to define precisely. Admittedly, we mostly use this level as a general category for higher-order cognitive functions that do not fit neatly into the perception, valuation, or execution levels. Nevertheless, a few studies have used fMRI or MEG to align language models with human semantic representations \citep{schwartz_inducing_2019, toneva_interpreting_2019}, including in multilingual contexts \citep{negi_brain-informed_2025}, or to improve the performance of visual models \citep{nishida_brain-mediated_2020} or auditory models \citep{moussa_improving_2025} on semantic tasks. These results, along with the representational similarity observed between language models and the human brain \citep{schrimpf_neural_2021, caucheteux_evidence_2023}, suggest that the brain regions and cognitive processes associated with the integration level could be promising targets to overcome the shallow conceptual integration observed in foundation models, at least if such methods can be scaled. Guiding models using neural signals from regions associated with social cognition or self-cognition could also be an interesting opportunity to explore, particularly in the context of AI agents. The relevance of this research direction could depend on the evolution of agentic systems, and particularly on the extent to which future agents will be expected to collaborate with each other and with human users. 

Altogether, the brain regions that we identified across the perception, valuation, execution, and integration levels are associated with important cognitive processes, and could allow us to monitor the evolution of meaningful neural signals over time. Nevertheless, this list should only be considered as a starting point, since many other brain regions and cognitive processes, as well as the functional connectivity or interactions between them, could provide valuable information for foundation models. Furthermore, while we largely focused on fMRI studies, where the link between regions and processes is relatively straightforward, we must remember that other neuroimaging techniques could offer complementary insights into human brain activity at different spatial and temporal scales. Interestingly, most of the research available at the perception and integration levels follows an approach that we could call \textit{pattern decoding}, where classical machine learning or deep learning models are aligned with neural representations in visual, auditory, or language regions. By contrast, the research available at the valuation level follows an approach that we could call \textit{cognitive and computational inference}, where reinforcement learning models are trained using neural signals reflecting specific cognitive and computational processes, such as reward and error signals extracted from EEG data. As we noted earlier, the transition to broad, open-ended foundation models could represent a cognitive turning point regarding the most promising neural signals to consider, since the current limitations of these models extend beyond the perception level. This turning point could make cognitive and computational inference increasingly important, as this approach could provide us with relevant higher-order neural signals, particularly for the valuation and execution levels. Importantly, available research demonstrates that human brain data can serve as a relevant training resource even when it is acquired from a limited number of participants, since model improvements have been observed with as few as six \citep{negi_brain-informed_2025} or eight \citep{moussa_improving_2025} human subjects. Furthermore, while the relatively low signal-to-noise ratio associated with neuroimaging data can be challenging, in particular when models are trained on individual subjects \citep{freteault_alignment_2025}, the effects of noise do not seem to represent an insurmountable obstacle, as they can be mitigated, for example, by performing sample selection \citep{fong_using_2018} in fMRI studies or electrode selection \citep{hansen_brain-guided_2025} in EEG studies, alongside a series of classical preprocessing methods such as spatial filtering \citep{kim_intrinsic_2017}. 

We introduced this paper by suggesting that a diversity of neural signals could potentially be leveraged to gain a better understanding of the cognitive processes that determine, for example, the choice of a particular word. Based on the promising brain regions and cognitive processes that we identified, we can now elaborate on this example. Let us imagine that a human subject is choosing their next written or spoken word, while their brain activity is recorded using one or several neuroimaging techniques. At the valuation level, neural signals from the ventral striatum could reveal the expected reward associated with the candidate word, while the activity of the vmPFC and OFC could reflect the perceived value of the current text or speech, or the confidence in the current verbal reasoning. Other valuation markers could account for the novelty or salience of the word, or for the effort or self-control involved. At the execution level, neural signals from the DLPFC could uncover the cognitive control mechanisms associated with the word choice, and reveal whether this decision is predominantly driven by stimuli, context, or instructions. The activity of the ACC could reflect the opportunity to change the current verbal reasoning, with the FPC encoding the reliabilities of alternative words or sentences, while other execution markers could account for motivation and attention. Finally, at the integration level, neural signals from the TPJ, mPFC, and PCC could open a window into the social or subjective dimensions behind the candidate word. Overall, whereas human-generated data is limited to observable actions, brain-generated data could give us access to values, motivations, hesitations, efforts, influences, alternatives, and other hidden variables of the cognitive latent space, in line with our hypothesis that $B^*(t)$ could include elements of $B(t)$ that are not included in $A(t)$. Nevertheless, leveraging these insights for foundation models would require dedicated methods to prioritize the use of limited neuroimaging data for strategically chosen, high-value steps in foundation model training. In the next section, we propose two such methods, discuss temporal scale synchronization and scaling, and suggest that in the long term, brain-trained foundation models could open new horizons for agents, AGI, and ASI. 

\section{Strategies for Brain-Trained Foundation Models}

\begin{figure}[h]
\begin{center}
\includegraphics[width=17cm]{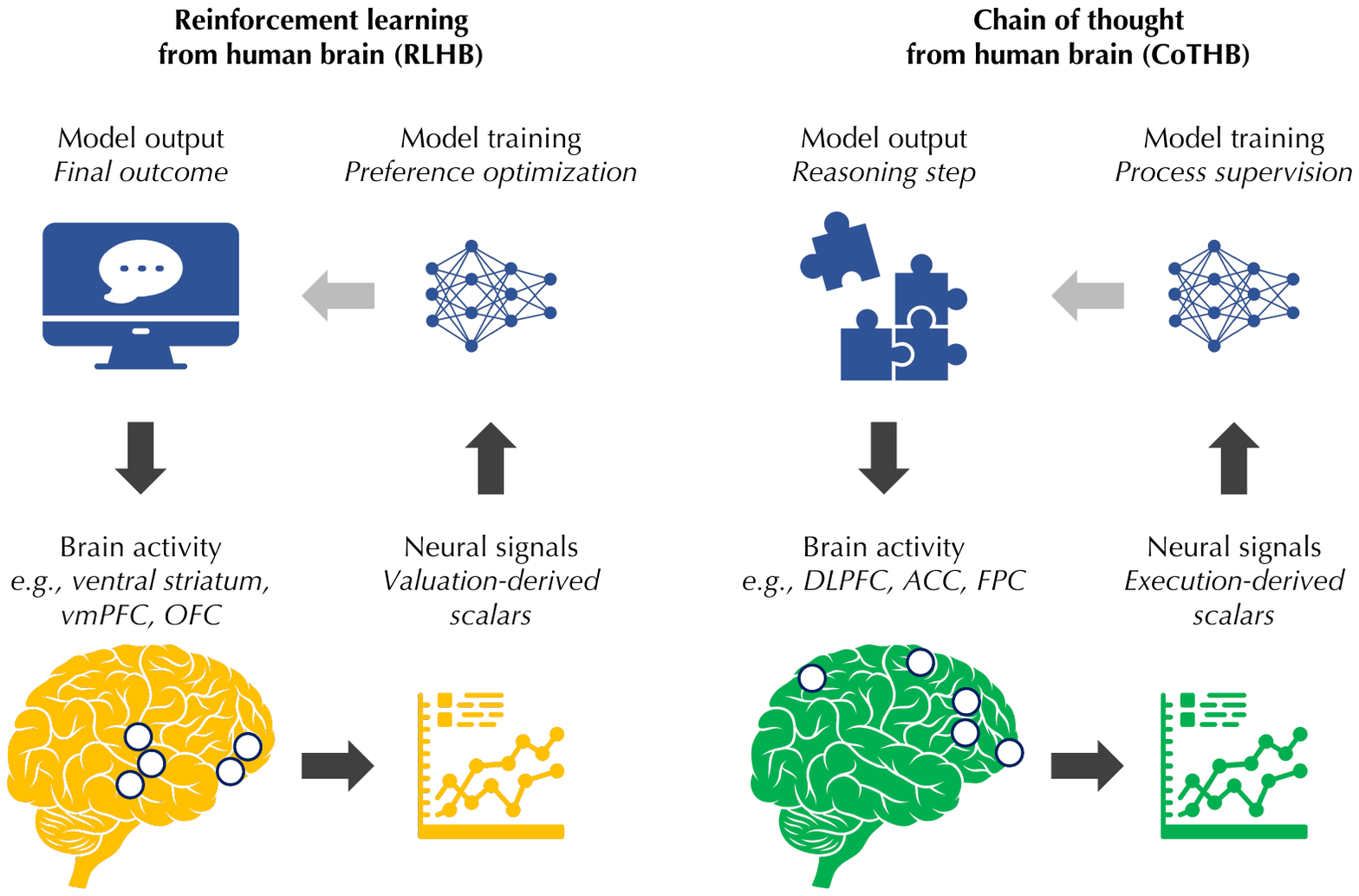}
\end{center}
\caption{\textbf{RLHB and CoTHB.} \textit{Left:} In RLHB, a final model output is presented to a subject, brain activity is recorded (here from valuation regions) as the subject evaluates the output, and neural signals are extracted to align the model. The grey arrow suggests that future outputs could be improved based on this feedback. \textit{Right:} In CoTHB, a reasoning step is presented to a subject, brain activity is recorded (here from execution regions) as the subject engages in reasoning, and neural signals are extracted to guide the model. The grey arrow suggests that the model could subsequently improve its ability to solve similar reasoning problems.}\label{fig:rlhb_and_cothb}
\end{figure}

Brain-trained foundation models would require dedicated methods to incorporate neural signals into their training strategies. While discoveries in cognitive and computational neuroscience have uncovered many promising brain regions and cognitive processes, existing research on guiding classical machine learning, deep learning, or reinforcement learning models using neuroimaging data has been relatively modest in scale, and focused on a narrow set of tasks. This limited scope was most certainly necessary, considering the heterogeneity and cost of neuroimaging data. However, foundation models have, by contrast, emerged as powerful systems trained on large-scale data and deployed across a wide range of applications. Therefore, incorporating neural signals into their training strategies could require either scaling current efforts significantly, or prioritizing the use of limited neuroimaging data for strategically chosen, high-value steps in foundation model training. As a basis for reflection, let us consider the proposition that leveraging human brain data could be useful in two scenarios: either when its acquisition becomes less expensive than behavioral data, or when the information of interest cannot be elicited through behavior \citep{mineault_neuroai_2025}. For the perception level and the linguistic tasks within the integration level, where existing research follows the pattern decoding approach, scaling current efforts could represent a reasonable path forward. Indeed, the alignment of visual, auditory, or language models with neural representations typically involves relatively large brain regions, which can increase the amount of information obtained per neuroimaging acquisition, thereby lowering the cost. By contrast, for the valuation and execution levels, where cognitive and computational inference has the potential to become increasingly relevant, prioritizing the use of neuroimaging data for strategically chosen training steps could be a strong alternative. Here, we propose two general methods for this second objective, which extend, respectively, the RLHF and CoT approaches to incorporate human brain data, as illustrated in Figure~\ref{fig:rlhb_and_cothb}. Importantly, the objective of this theoretical paper is not to demonstrate the effectiveness of specific algorithms, as such demonstration would require a different, experimental research approach. Therefore, the two methods we propose should only be understood as a high-level framework, mapping particular neural signals from the human brain with particularly strategic steps in foundation model training. Still, for both methods, we suggest several possible algorithmic paths and one concrete mathematical formulation that might be considered for leveraging the neural signals of interest, with the objective of improving, respectively, the alignment and reasoning capabilities of foundation models. 

\subsection{Reinforcement Learning from Human Brain}

The first method could be called \textit{reinforcement learning from human brain (RLHB)}. It would be an extension of RLHF \citep{christiano_deep_2017} incorporating neuroimaging data, and possibly other real-time measures, for a better alignment with human values. Integrating such a method into the training strategies of foundation models seems realistic, both from a technological and a human perspective. For example, adding EEG recording to an RLHF session should only increase the cost marginally, and assuming that appropriate EEG headsets are used, it should not represent a significant problem for the comfort of the human annotators. Adding fMRI recording would be more expensive and likely result in more discomfort, since the annotators would need to remain immobile inside an fMRI scanner, but the neural signals extracted would be more comprehensive. For both neuroimaging modalities, the cost and complexity, as well as the quality of the neural signals acquired, would naturally depend on the exact acquisition protocol involved, for example the type and number of EEG electrodes. Critically, RLHF is already an expensive and time-consuming post-training step \citep{xu_rlthf_2025}, and it is increasingly considered a bottleneck in the training strategies of foundation models \citep{yuan_self-rewarding_2025}, suggesting that neuroimaging could be a feasible and valuable addition in this context. It is also a step requiring a significant cognitive effort from the annotators, suggesting that the relative discomfort introduced by the neuroimaging measures could be compensated if it lowers the constraints regarding the number, precision, or comprehensiveness of the ratings required. Alongside neuroimaging data, RLHB could also benefit from other real-time measures, such as eye movements, heart rate, skin conductance, or facial muscle activity, which are non-invasive and may provide valuable insights into the degree of well-being, surprise, stress, attention, conflict, or cognitive effort experienced by the annotators. Interestingly, eye tracking could be used to align neural signals with the successive segments of a foundation model output, offering the possibility to train a fine-grained reward function, in line with current research on fine-grained human feedback \citep{wu_fine-grained_2023}. The ethical and social considerations, in particular the trade-off between the additional protections required for the annotators and the benefits expected from a better alignment of foundation models, will be addressed in the next section. 

A diversity of brain regions and cognitive processes could potentially be targeted for RLHB. At the valuation level, the monitoring of rewards, values, and confidence could effectively complement ratings as additional expressions of human preferences. The neural signals associated with novelty and salience may be useful to detect unexpected outputs, while the markers of effort, self-control, and affective processes may reveal the difficulty or ambiguity of the evaluation. These indicators could be extracted from the activity of regions such as the ventral striatum, vmPFC, OFC, amygdala, and insula. At the execution level, the monitoring of rules, alternatives, and the environment could be useful when the output is complex to evaluate, or when the annotators are asked to compare several possible outputs. The neural signals associated with motivation, attention, and action selection may provide an additional measure of their decision-making process. These indicators could be extracted from the activity of regions such as the DLPFC, ACC, FPC, PPC, and pre-SMA. Finally, at the integration level, the markers associated with conceptual integration, social cognition, and self-cognition may also be relevant, particularly for outputs with a strong intellectual, collective, or introspective dimension. These indicators could be extracted from the activity of the TPJ, mPFC, PCC, and related regions. Regarding the specific algorithms that could be used to extend RLHF into RLHB, we can think of several possibilities. Reward or value signals could be used as direct expressions of preference alongside the classical reward function, whereas negative affective markers could serve as regularizers for penalizing some outputs. Confidence signals could be used to weight the ratings according to their reliability, whereas novelty or salience signals could be leveraged to prioritize the evaluation on unexpected or intriguing outputs. Alternatively or concurrently, the reward function could be trained to predict neural signals alongside explicit ratings, in order to increase its robustness. Furthermore, it is worth noting that the continuous nature of human brain data may provide, in some cases, the opportunity for a smoother fine-tuning compared to discrete ratings. Overall, while the specific neural signals and algorithms would likely depend on each training strategy, RLHB could emerge as a promising method for improving the alignment of foundation models with human values. For a more concrete illustration of how RLHB could work, let us consider the following equation, which defines a policy optimization objective integrating valuation signals:

\begin{equation}
\begin{aligned}
\label{eq:rlhb}
J(\theta) = \mathbb{E}_{x \sim D,\; y \sim \pi_{\theta}(\cdot \mid x)} \Big[
& r_{\phi}(x, y)
+ \alpha \, B^{\mathrm{valuation}}(x, y) - \beta \log \frac{\pi_{\theta}(y \mid x)}{\pi_{\mathrm{ref}}(y \mid x)}
\Big]
\end{aligned}
\end{equation}

In this mathematical formulation, $J({\theta})$ represents the policy optimization objective, $\pi_{\theta}$ the current policy, and $\pi_{\mathrm{ref}}$ the reference policy. Similar to classical RLHF, the policy optimization relies on a reward model $r_{\phi}$ trained on expressed human preferences, with $D$ corresponding to the preference dataset, $x$ to the model input, and $y$ to the model output. The effect of this reward model is regularized by a Kullback-Leibler term, whose weight is controlled by a parameter $\beta$. The novelty resides in the addition of a new term: $B^{\mathrm{valuation}}$, a scalar derived from a valuation-related neural signal associated with each input-output pair, and whose weight is controlled by a parameter $\alpha$. This neural signal serves as an additional source of reward complementing the expression of human preferences, and could be extracted from brain regions such as the ventral striatum, vmPFC, or OFC. Importantly, this neural signal could either be directly measured, or predicted using a neural predictor model trained on available neuroimaging data. Indeed, while classical RLHF relies on human feedback, it must generalize from available examples during the policy optimization, in order to evaluate input-output pairs that have never been presented to humans. Similarly, while RLHB would rely on neuroimaging data, it would need to generalize from available recordings, in order to incorporate neural signals that have never been measured in human brains. Furthermore, we assume that the neural signal of interest would be appropriately preprocessed and normalized, so that a meaningful scalar can be derived from it. While using $B^{\mathrm{valuation}}$ as an additional source of reward could provide a new lever for optimizing foundation models based on human valuation processes, it should be noted that such an algorithm would use brain-generated data as a complement, rather than a replacement, for human-generated data. Very recently, EEG signals have been successfully used for aligning LLM outputs with human preferences in the context of image generation \citep{zhang_eeg-based_2026}, possibly providing foundational experimental evidence for leveraging neural signals in the context of alignment. 

\subsection{Chain of Thought from Human Brain}

The second method, perhaps more forward-looking, could be called \textit{chain of thought from human brain (CoTHB)}. It would be an extension of CoT \citep{wei_chain--thought_2023} incorporating neuroimaging data, and possibly other real-time measures, for a more trustworthy emulation of human executive function. Integrating such a method into the training strategies of foundation models could require the acquisition of new neuroimaging datasets focused on human reasoning, but these investments may be justified by the strategic importance of this research direction. Critically, frameworks designed to expand the flexibility of CoT often imply a greater computational cost \citep{yao_tree_2023, besta_graph_2024}, and still underperform direct answering across many model scales and benchmark complexities \citep{zheng_curse_2025}, suggesting that alternative paths such as neuroimaging could be worth considering for improving their performance. Fundamentally, CoTHB could focus on a subset of brain regions and cognitive processes particularly engaged in human reasoning, using neural signals such as the reliability of the current rule, the reliabilities of alternative rules, the value of the exploratory behavior, the retrieval of a previous rule, and the confirmation of a new rule. These indicators could be extracted, respectively, from the vmPFC, FPC, ACC, DLPFC, and ventral striatum. Regarding the specific algorithms that could be used to extend CoT into CoTHB, we can think of several possibilities. The current reasoning path could be followed as long as the reliability of the current rule is high, and alternative branches could be explored when the value of the exploratory behavior reaches a certain threshold. Alternatively or concurrently, neural signals could be used to drive the retrieval of a previous path, the confirmation of an exploratory path, or the monitoring of a specific number of alternative branches. As for RLHB, CoTHB could also benefit from other real-time measures alongside neuroimaging data. The amount of data that should be acquired to provide enough examples of human reasoning remains an open question, but it is reasonable to assume that some degree of generalization should be attainable from a limited number of carefully designed reasoning experiments \citep{donoso_toward_2026}. As for RLHB, the ethical and social considerations will be addressed in the next section. 

Compared with RLHB, using neuroimaging data in the context of CoTHB raises an interesting question. Human reasoning on complex problems can be long, often spanning several hours or even days, whereas neuroimaging data can only be acquired for shorter periods. Therefore, the insights gained from neural signals on relatively short reasoning tasks may, or may not, generalize well to the longer problems that we could expect foundation models to solve. This is less likely to be a difficulty for RLHB, where the evaluation of a foundation model output could be done in a shorter amount of time. One natural method to address this challenge in CoTHB could be cognitive compositionality: alternating between reasoning paths at different levels of abstraction, with high-level paths decomposing the problem into various steps, and low-level paths subsequently solving these steps. Here, neural signals associated with the value of the exploratory behavior, or with the reliabilities of alternative rules, could reflect the uncertainty associated with a particular step, and drive the creation of a dedicated low-level path to solve a specific issue. We could even imagine an algorithm integrating both branching and compositionality, which would continue the reasoning path if the reliability of the current rule is high; explore an alternative branch if the reliability of the current rule is low; and create a dedicated low-level path to reduce the uncertainty for a particular step if the reliability of the current rule is intermediate. Regardless of the specific algorithm, compositionality has already demonstrated its potential for foundation models \citep{zhou_least--most_2023}, and could be a promising approach to effectively leverage the insights of neuroimaging data for long reasoning problems. At a more fundamental level, this raises the question of whether the cognitive processes underlying human reasoning are close to an optimal computational trade-off \citep{koechlin_evolutionary_2014}, or if significantly more efficient reasoning models can be discovered. Although this remains an open question, training foundation models directly on human brain data could provide us with a starting point for experimenting on this important subject. Overall, while the specific neural signals and algorithms would likely, as for RLHB, depend on each training strategy, CoTHB could emerge as a promising method for refining the emulation of human executive function in foundation models. For a more concrete illustration of how CoTHB could work, let us consider the following equation, which defines a policy optimization objective integrating executive signals:

\begin{equation}
\begin{aligned}
\label{eq:cothb}
J(\theta) =
\mathbb{E}_{(x,h) \sim D,\; z \sim \pi_{\theta}(\cdot \mid x,h)}
\Big[
& r_{\psi}(x,h,z)
- \beta \log
\frac{\pi_{\theta}(z \mid x,h)}
{\pi_{\mathrm{ref}}(z \mid x,h)}
- \gamma \, B^{\mathrm{execution}}(x,h,z)
\Big]
\end{aligned}
\end{equation}

In this mathematical formulation, $J({\theta})$, $\pi_{\theta}$, $\pi_{\mathrm{ref}}$, and $x$ represent as before the policy optimization objective, the current policy, the reference policy, and the model input. However, instead of considering a model output $y$, we are now interested in $h$, the history of the reasoning sequence already generated, and $z$, the candidate continuation of this reasoning sequence. As a result, $D$ now corresponds to a process supervision dataset. The reward model $r_{\psi}(x,h,z)$, trained on expressed human choices, determines the preference for a given continuation at a particular step of the reasoning process, rather than the preference for a final model outcome. This mechanism aligns with current strategies for optimizing reasoning capabilities using process supervision algorithms \citep{uesato_solving_2022, lightman_lets_2023}. The novelty resides in the addition of a new regularization term: $B^{\mathrm{execution}}$, a scalar derived from an execution-related neural signal associated with each input-history-continuation triplet, and whose weight is controlled by a parameter $\gamma$. This neural signal serves as evidence that the candidate continuation is unreliable, and therefore that alternative reasoning paths could be considered. It could be extracted from brain regions such as the DLPFC, ACC, or FPC. As before, this neural signal could either be measured or predicted, and it is assumed to be appropriately preprocessed and normalized so that a meaningful scalar can be derived from it. Using $B^{\mathrm{execution}}$ as an additional regularization term could provide a new lever for optimizing foundation models based on human executive processes, and it could eventually serve as a foundation for more sophisticated algorithms. For example, if most continuation candidates for a given history are strongly regularized regardless of the expressed human choices, this could be interpreted as evidence that the current reasoning path is unreliable, and trigger a mechanism for backtracking to the previous step in the reasoning process. 

\subsection{Synchronizing Brain and Model Temporal Scales}

Whereas neuroimaging techniques such as EEG or MEG offer a high temporal resolution, the temporal resolution provided by other techniques such as fMRI is lower. For RLHB and CoTHB, this is unlikely to present a problem at inference time, because we are not proposing a closed-loop system where neuroimaging acquisition and token generation should directly synchronize. Rather, both methods assume an offline training process, where the alignment and reasoning capabilities of foundation models are optimized using neural signals before the brain-trained foundation models are deployed. By contrast, the question of temporal resolution becomes highly relevant at training time, where the temporal scale of neuroimaging acquisition determines the possible uses of the neural signals. Here, two situations could be distinguished. In their primary formulation, neither RLHB nor CoTHB necessarily require a high temporal resolution, because the subjects are expected to evaluate either a final model outcome or a reasoning continuation, two sequence types that are assumed to be substantially longer than a single word or token. In such a context, leveraging a slow neuroimaging technique such as fMRI seems plausible, since fMRI has already been used for investigating the valuation processes elicited by short texts \citep{bohrn_when_2013}, as well as the executive processes elicited by reasoning problems \citep{fangmeier_fmri_2006}. However, we can also envision a secondary, more advanced formulation where neural signals would be associated with specific segments of a foundation model output, as we suggested when discussing RLHB, or even associated with specific words or tokens, in order to train a more fine-grained reward function. For such an advanced method, the low temporal resolution of fMRI may indeed represent a challenge, and faster techniques such as EEG or MEG could be preferred. Existing research on guiding deep learning models for linguistic tasks using neuroimaging data is particularly informative in this respect. While MEG can be more directly leveraged to extract the neural signals associated with individual words, which can then be successfully used to guide model training \citep{schwartz_inducing_2019}, leveraging fMRI in this context is possible but more challenging, as the training method is constrained by the temporal delay and overlap between the neural signals \citep{negi_brain-informed_2025}. As new methods are developed for analyzing the neural correlates of written and spoken language at different temporal scales using fMRI \citep{chen_cortical_2024}, it is possible that this technique may eventually become more widely usable. Nevertheless, if advanced RLHB and CoTHB methods requiring fine-grained attribution are explored, or if even more advanced, closed-loop alternatives are envisioned, the safer bet would probably be to focus on techniques such as EEG or MEG, whose characteristics align more naturally with these particular objectives. 

\subsection{Scaling Brain-Trained Foundation Models}

In this theoretical paper, we do not assume that currently available neuroimaging datasets are necessarily sufficient for developing useful brain-trained foundation models. Rather, we expect that both RLHB and CoTHB would require dedicated experimental research and progressive scaling before they could potentially be deployed within the post-training pipelines of foundation models. This raises the complex scientific question of the size and diversity of the neuroimaging corpus that would be necessary for scaling brain-trained foundation models. How much neuroimaging data would we need for meaningfully improving the performance of large-scale models, all the while preventing overfitting on the acquired neuroimaging datasets? Scaling laws could be very different for human-generated data, which is typically expressed in symbolic language and may include social, cultural, or demographic biases \citep{ali_sociotechnical_2025}, compared with brain-generated data, which is usually noisy and subject to its own biases, but might provide a unique access to some universal mechanisms of human cognition. As a result, a direct comparison with the size and diversity of available classical datasets would only provide limited evidence on the potential of human brain data. It is also important to acknowledge that the value of neural signals and the risk of overfitting should not be assessed in isolation, since RLHF and SFT sessions are already dependent on the cognitive effort of relatively limited groups of human supervisors. Therefore, the question could really be formulated as the size and diversity of the neuroimaging corpus that would be necessary for meaningfully improving the performance achieved through current RLHF and SFT steps, all the while introducing the least possible additional overfitting risks to these steps. Similarly, the cost of RLHB, for example, should be evaluated as the supplementary expense incurred when adding neuroimaging acquisition to a RLHF session, rather than compared to the cost of a standalone neuroimaging session. Overall, the challenge could be defined as evaluating the expected performance gains, overfitting risks, and additional costs of using $B^*(t)$ to access elements of $B(t)$ that are not included in $A(t)$. Existing research on guiding classical machine learning, deep learning, or reinforcement learning models using neural signals provides important insights on the feasibility of the training process, and demonstrates that performance gains can be achieved despite the noise, limited number of subjects, and other challenging characteristics of neuroimaging datasets. However, beyond the fact that RLHB and CoTHB would require their own experimental validation, available results are insufficient to establish the scaling laws of brain-generated data, or to evaluate the risk of overfitting if this strategy was deployed at scale. This suggests the need for a preliminary scalability roadmap for brain-trained foundation models. 

An effective roadmap should probably start with currently available models and data, then consider the acquisition of focused neuroimaging datasets, and eventually, more conventional scaling efforts if early experimentation is successful. A particularly attractive starting point could be to explore the potential of brain prediction models \citep{adeli_predicting_2023} pre-trained on large neuroimaging datasets \citep{allen_massive_2022}. Such prediction models have already been used to guide deep learning models for visual tasks by generating large-scale, synthetic fMRI data without additional cost \citep{zhao_neuralood_2025}, suggesting their relevance as a baseline resource for brain-trained foundation models. Very recently, a foundation model of vision, audition, and language has been developed, capable of predicting the neural correlates associated with linguistic stimuli \citep{dascoli_foundation_2026}, which could make it potentially usable for RLHB and CoTHB experimentation. Beyond these particular examples, the large-scale models developed specifically for neuroscience, sometimes known as brain foundation models, could support the initial scaling stages on two complementary levels. First, the ecosystem established around brain foundation models, including open-access data repositories, open-source code, and standardization methods for various neuroimaging modalities \citep{tak_generalizable_2026, el_ouahidi_reve_2025} could be repurposed either to improve the generation of synthetic neuroimaging data, or even more directly, to develop the neural predictor models yielding the valuation-derived scalars and execution-derived scalars in RLHB and CoTHB. Second, if dedicated neuroimaging datasets are acquired to develop the neural predictor models, fine-tuning relevant brain foundation models could potentially emerge as a more effective method than training these predictors from scratch, as transfer learning for neuroimaging analysis can often improve performance and spare computational resources \citep{ardalan_transfer_2022}. Overall, brain foundation models and their ecosystem could allow us to take advantage of the full extent of relevant neuroimaging data, thereby reducing our dependence on the specific biases or sources of noise of small-scale datasets, and potentially mitigating the risk of overfitting. However, while pre-trained models could represent a valuable resource for conducting early experiments and increasing robustness in a cost-effective manner, they would still be limited by the scope and characteristics of their own training corpus. In particular, their training data may not cover all the neural signals relevant for valuation or execution, limiting their ability to generalize to applications such as RLHB or CoTHB. This suggests that at a certain point in the roadmap, dedicated neuroimaging datasets may need to be acquired. 

Once this stage is reached, two complementary experimental directions could be prioritized to mitigate specific overfitting risks. First, experiments could be conducted in contexts where neuroimaging acquisition is particularly practical and relatively inexpensive, for example by adding simple EEG recording devices to already planned RLHF sessions. This could facilitate the acquisition of neuroimaging data on a reasonable number of participants, and help prevent overfitting on individual brain patterns. Second, if more sophisticated experiments are conducted, multimodal experiments combining several neuroimaging techniques, for example fMRI and EEG, could be prioritized in order to diversify the neural signals and help prevent overfitting on the particularities of each neuroimaging modality. Of particular interest here, a multimodal brain foundation model for fMRI, EEG, and MEG has very recently been developed, able to capture modality-invariant representations \citep{guo_brain-_2026}, which suggests that brain foundation models may be helpful for transfer learning in multimodal contexts as well. Another specific strategy to prevent overfitting could be to augment neuroimaging data by applying physiologically meaningful perturbations to acquired datasets \citep{zlatov_towards_2022}, potentially improving robustness. If such experiments are successful, more conventional scaling methods could then be envisioned, with an increase in the number and diversity of participants, an appropriate normalization across subjects, sessions, and modalities, and a deeper integration into the post-training pipelines of foundation models. Importantly, each stage of this preliminary scalability roadmap corresponds to a particular generalization hypothesis, as the objective of the proposed steps is to evaluate whether the predictions of brain foundation models generalize to valuation or execution tasks; whether the neural signals provide consistent information on valuation or execution processes across subjects and modalities; and ultimately, whether this information is worth incorporating into foundation models. Regardless of the specific trajectory that may be considered for scaling brain-trained foundation models, neural signal prediction, transfer learning, multimodal acquisition, and data augmentation could represent promising strategies to explore for preventing overfitting. 

\subsection{Toward Brain-Trained Agents}

Neural patterns from the perception level can be leveraged to guide visual or auditory models. Neural signals from the valuation level could be particularly relevant for RLHB, and neural signals from the execution level for CoTHB. However, the case of the integration level is more intriguing. While its neural patterns can be used to guide language models, we speculate that markers associated with conceptual integration, social cognition, and self-cognition could eventually be leveraged for the development of agents. The evolution of agentic systems is difficult to predict, but it seems reasonable to assume that future agents will be expected to collaborate more thoroughly with each other \citep{park_generative_2023} and with human users \citep{afzoon_modeling_2025}, in order to perform increasingly complex tasks in their environment \citep{wang_voyager_2023}. Recent research has explored cognitively inspired architectures allowing agents to generate and evaluate hypotheses about other agents \citep{cross_hypothetical_2024}, as well as biologically inspired architectures allowing agents to perceive and interact with real-world environments \citep{liu_neural_2025}. The fact that neuroscience discoveries serve as a natural inspiration suggests that the incorporation of neuroimaging data may provide valuable insights for agentic systems, potentially opening a path toward brain-trained agents. 

\subsection{Toward Artificial General Intelligence}

Foundation models have already reached human-level performance on a range of professional and academic benchmarks \citep{openai_gpt-4_2024}. As a result, they are increasingly seen as strong candidates for reaching the hypothetical stage of AGI \citep{fei_towards_2022}, and may even already exhibit some characteristics of AGI \citep{bubeck_sparks_2023}, at least if we consider this concept in a functional sense. From our perspective, structural AGI would imply replicating the structure of the human brain, or discovering equivalent architectures, so that an AI system could be able to acquire its own knowledge and develop cognitive abilities up to the human level. By contrast, functional AGI would more narrowly imply replicating the functions of the human brain across a wide range of tasks, up to the level of knowledgeable individuals. While the latter objective seems more easily attainable, the current limitations of foundation models still prevent them from achieving the degree of generality and adaptability associated with human cognition \citep{raman_navigating_2025}. By leveraging brain-generated data alongside human-generated data, brain-trained foundation models could represent a distinctive strategy for attenuating some of the failure modes of current AI systems in perception, valuation, execution, and integration, bringing them arguably closer to functional AGI. There is, of course, no guarantee that neural signals would be sufficient to overcome all the limitations that stand between current foundation models and human-level intelligence, in particular because human brain data $B^*(t)$ is only an approximation of human brain activity $B(t)$. Nevertheless, there is a reasonable guarantee that all human cognitive functions that current foundation models are unable to approximate are implemented somewhere within the human brain, making neural signals, if not a certain bet, at least a natural path to explore. 

\subsection{Toward Artificial Superintelligence}

Looking further ahead, brain-trained foundation models might even open a path beyond functional AGI, toward the even more hypothetical stage of functional or structural ASI. If current foundation model architectures prove insufficient for integrating the insights of neural signals, these structural limitations could motivate the development of new architectures, more compatible with our knowledge of the human brain. Eventually, large-scale neuroimaging corpora could emerge as a uniquely informative benchmark for evaluating the biological plausibility, neural similarity, and cognitive alignment of foundation models, providing us with a starting point for experimenting and possibly developing systems beyond human abilities. This strategy would be compatible with the exploration of entirely new frameworks inspired by the human brain \citep{lecun_path_2022}, and in fact, it could represent a realistic and potentially effective middle ground between continuing to scale current architectures and exploring alternative, neuroscience-inspired solutions. By starting from well-established and widely used foundation models, and progressively adapting them so that they can better accommodate the insights of neural signals, we might be able to advance toward ASI in a robust and gradual manner, regardless of whether ASI can ultimately be considered as a well-defined stage in AI development, or is simply destined to serve as a theoretical upper bound for AI research. Arguably, such a process would share some similarities with the evolution of the human brain itself, where all the intermediate steps had to be functional and build upon each other in a natural progression. Any reflection on brain-trained agents, AGI, or ASI is necessarily speculative at this stage, but the long-term potential of brain-trained foundation models at these three successive levels of anticipation is worth considering nonetheless. Indeed, these three horizons arguably map to three hypothetical future developments in the use of human brain data for foundation model training: the exploration of new neural signals of interest, the exploitation of these signals to address new limitations in AI systems, and the development of new AI architectures more compatible with these neural signals. 

Theoretically, the promising brain regions and cognitive processes that we identified in this paper, as well as the general methods that we proposed, could be relevant for foundation models even if neuroimaging technologies remain at their current level. Practically, of course, this assumption is likely to be too conservative. Rather, it seems reasonable to expect that eventually, the advances in neural translation from one neuroimaging modality to another, as well as the generalization of neural interfaces, for example, could significantly increase the quality and volume of neuroimaging data at our disposal. While RLHB and CoTHB both eventually assume the acquisition of neuroimaging datasets specifically dedicated to foundation model training, a wider adoption of wearable or implantable neural technologies, for example in the context of games, could also provide us with large-scale human brain data that may be repurposed for this objective. Open-ended and complex games, in particular, could be promising targets \citep{mineault_neuroai_2025}. Interestingly, neural interfaces raise the possibility of adapting the outputs of foundation models to the neural signals of the users, searching for example for correlates of attention, curiosity, stress, fatigue, motivation, or effort at inference time, and modifying the length, complexity, or tone of the answers accordingly. Looking forward, this suggests that the interactions with foundation models could also become an opportunity for acquiring relevant neuroimaging data, monitoring the reactions of the users to specific answers, in a sort of neural feedback loop. Overall, while current foundation models have emerged from the vast corpus of human-generated data, reflecting the written and oral knowledge of humankind, brain-trained foundation models could potentially leverage the growing corpus of brain-generated data, reflecting the generative processes of human cognition that built this knowledge in the first place. There is some ambition in this strategy, which could bring us closer to combining the wisdom acquired over thousands of years by human history with the information accumulated over billions of years by biological evolution. Nevertheless, while these new horizons could provide extraordinary possibilities, they also raise important ethical, social, and technical questions, which we address in the next section. 

\section{Challenges and Opportunities}

Classical foundation models already come with significant ethical and social challenges. Their training data can contain personal or sensitive information acquired without explicit consent for this usage, and identities can be revealed despite the anonymization efforts. The training corpus can also include social, cultural, or demographic biases, and overrepresent particular groups or ideas. Brain-trained foundation models could introduce an additional layer of complexity, considering both the highly personal and sensitive nature of human brain activity, and the comparatively smaller number of individuals for whom neuroimaging data is currently available. Therefore, these models would almost certainly require stronger informed consent procedures, along with additional protections and clarifications regarding the privacy, security, anonymity, and ownership of the data. While many of these requirements already exist for neuroimaging datasets, the acquisition of such datasets has been, until now, primarily intended for scientific or medical research, suggesting that a new evaluation of the risks should be conducted if human brain data is also leveraged for foundation model training. Critically, whereas human-generated data is the result of human intentionality and decision-making, brain-generated data can be seen as an automatic and unfiltered product of human cognition. The fact that this data precedes and could potentially predict voluntary actions highlights the necessity of implementing strong measures to protect the freedom of thought in its most fundamental sense. We speculate that many of these questions would emerge anyway, and require adequate governance mechanisms, if neural interfaces become more widely used in the future. While these challenges are significant, they should be considered in light of the equally significant opportunities offered by brain-trained foundation models. In particular, a better alignment with human values, and a more trustworthy emulation of human executive function, could substantially improve the reliability of foundation models, expanding their human and social benefits. Furthermore, while the cost of neuroimaging data could concentrate its acquisition within economically developed regions, this equity challenge could be partially counterbalanced if grounding foundation models in the universal mechanisms of human cognition helps reduce the social, cultural, or demographic biases associated with linguistic contents. 

We already mentioned some of the specific technical challenges associated with human brain data. While neuroimaging data can be noisy, distinctively individual, and difficult to interpret, methods can be developed to address these difficulties. Meanwhile, the heterogeneity and cost of neuroimaging datasets are increasingly mitigated by large-scale repositories and standardization, and could be further overcome by potential advances in neural translation and neural interfaces. Ultimately, the best proof of concept for brain-trained foundation models could be the remarkable diversity of the discoveries in cognitive and computational neuroscience: if a variety of statistically significant effects can be uncovered and replicated despite the challenging nature of neuroimaging data, this suggests that many neural signals directly useful for foundation models could potentially be extracted as well. Interestingly, brain-trained foundation models could even represent a more straightforward application of neuroimaging data than many possible use cases of neural interfaces at an individual level. Indeed, whereas individual-level applications would require neural signals to be precise enough in order to deliver relevant feedback to each user, the training strategies of foundation models may more easily leverage group-level statistics, tolerating a higher threshold of variability and noise. Overall, many of the technical challenges associated with neuroimaging data reflect its singular nature: whereas human-generated data leverages a shared symbolic system, brain-generated data approximates the neural complexity of each individual, reflecting the particular ways in which each human brain processes experiences and thoughts. Since the interpretation of such neural complexity depends on the discoveries in cognitive and computational neuroscience, neuroimaging data is also fundamentally open-ended, in the sense that future discoveries could increase the value of present datasets by retroactively unlocking new neural signals of interest. Although such singularity and open-endedness strengthen the need for appropriate ethical and social measures, this evolving potential could offer significant opportunities in the long term, positioning neuroimaging data as a uniquely promising asset for the future of foundation models. 

\section{Conclusion}

\begin{figure}[h]
\begin{center}
\includegraphics[width=17cm]{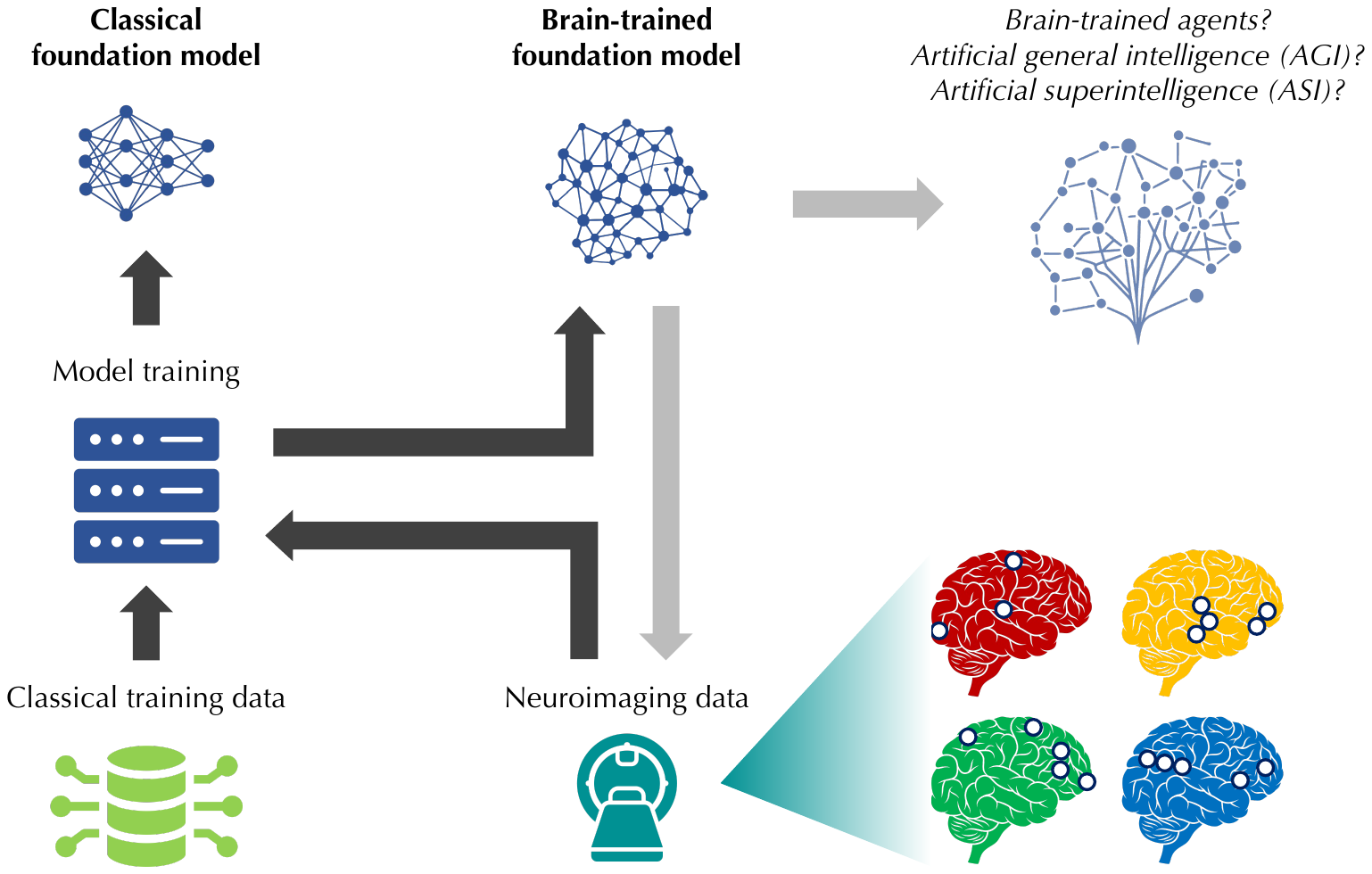}
\end{center}
\caption{\textbf{Future of foundation models.} \textit{Left:} In the current strategy, classical foundation models are trained on classical training data, represented in yellow-green to emphasize the importance of observable human actions, $A(t)$. \textit{Middle:} In the new strategy, brain-trained foundation models would incorporate neuroimaging data, $B^*(t)$, alongside classical training data. The vertical grey arrow suggests that this strategy could also motivate further advances in neuroscience, unlocking a virtuous circle. \textit{Bottom-right:} The promising brain regions and cognitive processes identified in this paper could help overcome the current limitations of classical foundation models. \textit{Top-right:} The horizontal grey arrow suggests potential future advances toward brain-trained agents, AGI, and ASI.}\label{fig:future_of_foundation_models}
\end{figure}

Foundation models have emerged as the dominant paradigm in contemporary AI. However, their reliance on observable human actions, $A(t)$, constrains them to surface-level regularities, which could partially explain some of their current limitations. In this paper, we explore a new strategy for AI: training foundation models directly on human brain data, $B^*(t)$, in order to approximate the generative processes of human brain activity, $B(t)$. We assume that neuroimaging data could give us access to a cognitive latent space that partially includes, but vastly exceeds, the narrow bandwidth of observable behavior. In other words, we hypothesize that $B^*(t)$ could include elements of $B(t)$ that are not included in $A(t)$. Alongside device-generated and human-generated data, we suggest that neuroimaging data could be distinguished as a new category, which we could call brain-generated data, in the context of foundation model training. We highlight a series of brain regions and cognitive processes whose neural signals could help overcome some of the current limitations of foundation models at the perception, valuation, execution, and integration levels. We propose two general methods, RLHB and CoTHB, that could be implemented to prioritize the use of limited neuroimaging data for strategically chosen, high-value steps in foundation model training, and discuss temporal scale synchronization and scaling, as well as the potential implications for agents, AGI, and ASI. We conclude by highlighting the ethical and social challenges of brain-trained foundation models, as well as the technical challenges associated with neuroimaging data, while also mentioning significant opportunities, such as a better and more universal alignment with human values. Importantly, brain-trained foundation models could represent a realistic and potentially effective middle ground between continuing to scale current architectures and exploring alternative, neuroscience-inspired solutions. The open-ended nature of neuroimaging data, as well as the potential advances in neural translation and neural interfaces, could make this strategy increasingly relevant over time, and unlock a virtuous circle at the intersection of neuroscience and AI, as illustrated in Figure~\ref{fig:future_of_foundation_models}. For the future of foundation models, perhaps we should start building neuroimaging platforms alongside data centers. 

\bibliographystyle{Frontiers-Harvard}
\bibliography{references}

\end{document}